\documentclass[graybox]{svmult}

\newcommand{\beq}{\begin{equation}}
\newcommand{\eeq}{\end{equation}}
\newcommand{\bea}{\begin{eqnarray}}
\newcommand{\eea}{\end{eqnarray}}

\newcommand{\Mpl}{M_{\rm P}}
\newcommand{\Madm}{M_{\rm ADM}}
\newcommand{\ellp}{R_{\rm P}}

\newcommand{\RC}{R_{\rm C}}

\newcommand{\tpl}{t_{\rm P}}

\usepackage{mathptmx}       
\usepackage{helvet}         
\usepackage{courier}        
\usepackage{type1cm}        
\usepackage{makeidx}         
\usepackage{graphicx}        
\usepackage{multicol}        
\usepackage[bottom]{footmisc}

\makeindex             

\begin{document}

\title*{Quantum Black Holes as the Link Between Microphysics and Macrophysics}
\author{B. J. Carrr}
\institute{B. J. Carr \at School of Physics and Astronomy, Queen Mary University  of London, Mile End Road, London E1 4NS, UK, \email{B.J.Carr@qmul.ac.uk}}
%
%
\maketitle

\abstract*{There appears to be a duality between elementary particles, which span the mass range below the Planck scale, and black holes, which span the mass range range above it. In particular, the Black Hole Uncertainty Principle Correspondence posits a smooth transition between the Compton and Schwarzschild scales as a function of mass. This suggests that all black holes are in some sense quantum, that  elementary particles can be interpreted as sub-Planckian black holes, and that there is a subtle connection between quantum and classical physics.
}  

\abstract{There appears to be a duality between elementary particles, which span the mass range below the Planck scale, and black holes, which span the mass range range above it. In particular, the Black Hole Uncertainty Principle Correspondence posits a smooth transition between the Compton and Schwarzschild scales as a function of mass.  This suggests that all black holes are in some sense quantum, that  elementary particles can be interpreted as sub-Planckian black holes,  and that there is a subtle connection between quantum and classical physics.
}

\section{Classical versus quantum black holes}
\label{sec:1}

At the previous Karl Schwarzschild meeting, I spoke about some quantum aspects of primordial black holes  \cite{Ca:2014a} and what I term the Black Hole Uncertainty Principle correspondence \cite{Ca:2014b}. My contribution this year will involve an amalgamation of these two ideas and is therefore a natural follow-up. It will also allow me to discuss some recent work with two of the organisers of this meeting! 

Black holes could exist over a wide range of mass scales. Those larger than several solar masses would form at the endpoint of evolution of ordinary stars and there should be billions of these even in the disc of our own galaxy. ``Intermediate Mass Black Holes'' (IMBHs) would  derive from stars bigger than $100~M_{\odot}$, which are radiation-dominated and collapse due to an instability during oxygen-burning, and the first primordial stars may have been in this  range.   
``Supermassive Black Holes'' (SMBHs), with masses from $10^6~M_{\odot}$ to $10^{10}~M_{\odot}$, are thought to reside in galactic nuclei, with 
our own galaxy harbouring one of $4 \times 10^6~M_{\odot}$
and quasars
being powered by ones of around $10^8~M_{\odot}$.  
All these black holes might be described as ``macroscopic'' since they are larger than a kilometre in radius. 

Black holes smaller than a solar mass could have formed in the early universe, the density being $\rho \sim 1/(Gt^2)$ at a time $t$ after the Big Bang. Since a region of mass $M$ requires a density $\rho \sim c^6/(G^3M^2)$ to form an event horizon, such ``Primordial Black Holes'' (PBHs) 
would  initially have of order the horizon mass $M_H \sim c^3t/G$,
so those forming at the Planck time ($t_P \sim 10^{-43}$s) would have the Planck mass ($M_P \sim 10^{-5}$g), while those forming at $t \sim 1$s 
would have a mass of $10^5 M_{\odot}$. Therefore PBHs could span an enormous mass range.  
Those initially lighter than $M_* \sim 10^{15}$ g 
would be  smaller than a proton and have evaporated by now due to Hawking radiation, the temperature and evaporation time of a black hole of mass $M$ being $T \sim 10^{12}(M/10^{15}g)^{-1}$K and $\tau \sim 10^{10}(M/10^{15}g)^{3}$y, respectively \cite{hawking}.
I will classify black holes smaller than $M_*$  as ``quantum'',
although I will argue later that {\it all} black holes are in a sense quantum. Those smaller than a lunar mass, $10^{24}$g, will be classified as ``microscopic'', since their size is less than a micron. Coincidentally, this is also the mass above which $T$ falls below the CMB temperature.

A theory of quantum gravity would be required to understand 
the evaporation process as the black hole mass falls to $M_P$ and this might even allow stable Planck-mass relics. The existence of extra spatial dimensions,
beyond the three macroscopic ones, may also come into play.
These dimensions are usually assumed to be compactified on the Planck length ($R_P \sim 10^{-33}$cm) but they can be much larger than this in some models. This would imply
that gravity
grows more strongly at short distances than implied by the inverse-square law \cite{arkani}, 
leading to the possibility of TeV quantum gravity and 
black hole production at accelerators.
Such holes are not themselves primordial but this would have crucial implications for PBH formation. 

\begin{figure}[b]
\sidecaption
\includegraphics[scale=.5]{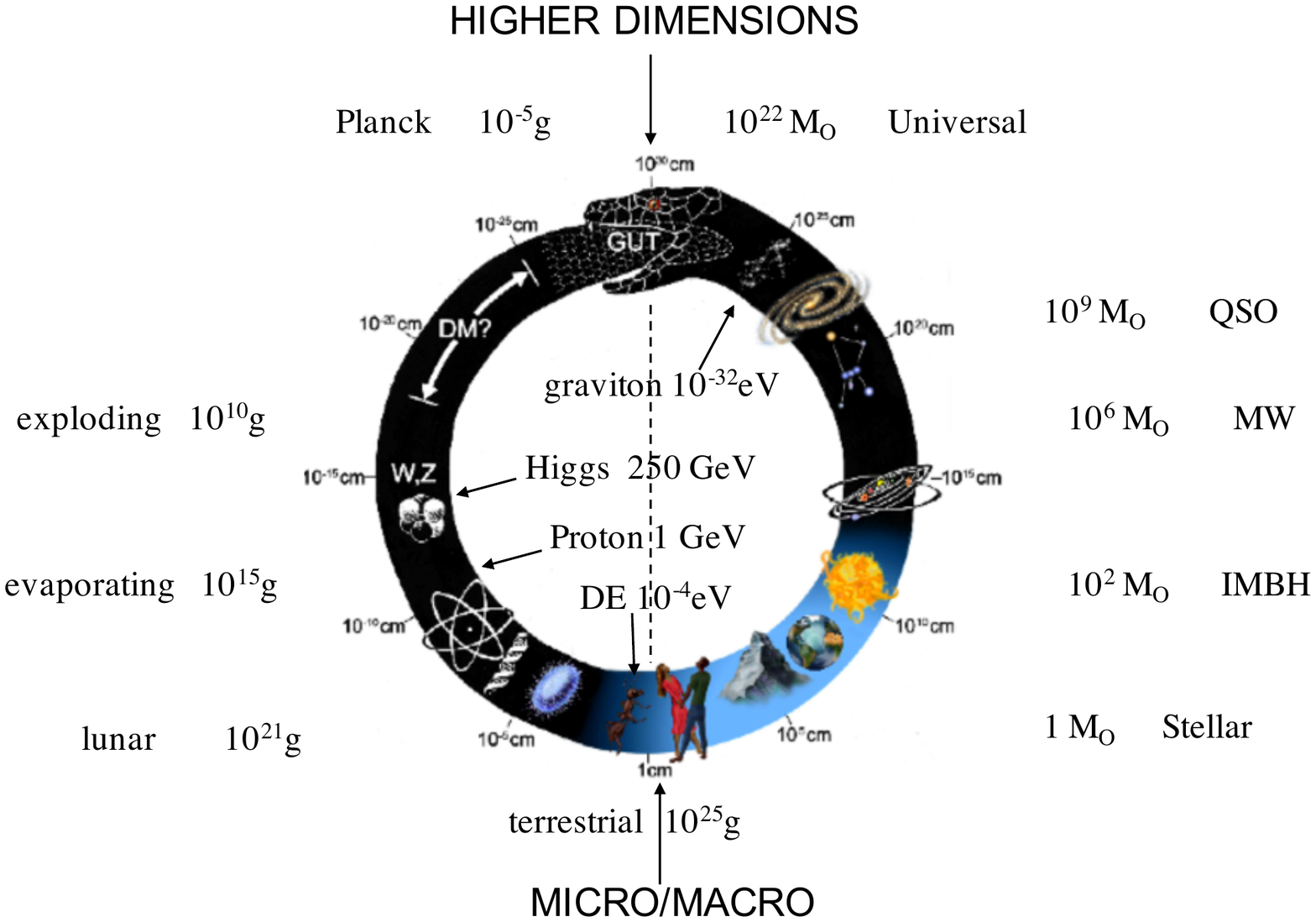}
\caption{
The Cosmic Uroboros is used to indicate that mass the various  types of black holes and elementary particles, the divison between the micro and macro domains being indicated by the vertical line. QSO stands for ``Quasi-Stellar Object'', MW for ``Milky Way'', IMBH for ``Intermediate Mass Black Hole'', LHC for ``Large Hadron Collider'', and ``DE'' for "Dark Energy". 
\label{fig:urob}
}
\end{figure}

The wide range of masses of black holes and their crucial role  in linking macrophysics and microphysics is summarized in Fig.~\ref{fig:urob}. This shows the Cosmic Uroborus (the snake eating its own tail), with the  various scales of structure in the universe indicated along the side. It can be regarded as a sort of ``clock'' in which the scale changes by a factor of $10$ for each minute -- from the Planck scale
at the top left to the scale of the observable universe 
at the top right. 
The head meets the tail at the Big Bang because at the horizon distance one is peering back to an epoch when the universe was very small, so the very large meets the very small there. 
The various types of black holes discussed above are indicated on the outside of the Urobrous.
They are labelled by their mass, this being proportional to their size if there are three spatial dimensions. On the right are the well established astrophysical black holes.
On the left -- and possibly extending somewhat to the right -- are the more speculative PBHs.
The vertical line between the bottom of the Uroborus (planetary mass black holes) and the top (Planck mass black holes and extra dimensions) provides a convenient division between the microphysical and macrophysical domains. 

Although the length-scale $\lambda$ decreases as one approaches the top of the Uroborus from the left, the mass of the associated particle $m \sim \hbar/(\lambda c)$ increases. So Fig.~\ref{fig:urob} can also be used to represent elementary particles. 
On the inside of the Uroborus are indicated the positions of the Higgs boson ($250$~GeV) and proton ($1$~GeV) on the left, the dark energy mass-scale ($10^{-4}$eV) at the bottom, and the mass (limit) on the gravitino ($10^{-32}$eV) at the top. 
Note that the inner scale also gives the temperature of a black hole with mass indicated by the outer scale.

\section{The Black Hole Uncertainty Principle Correspondence}
\label{sec:2}

\begin{figure}[b]
\sidecaption
\includegraphics[scale=.35]{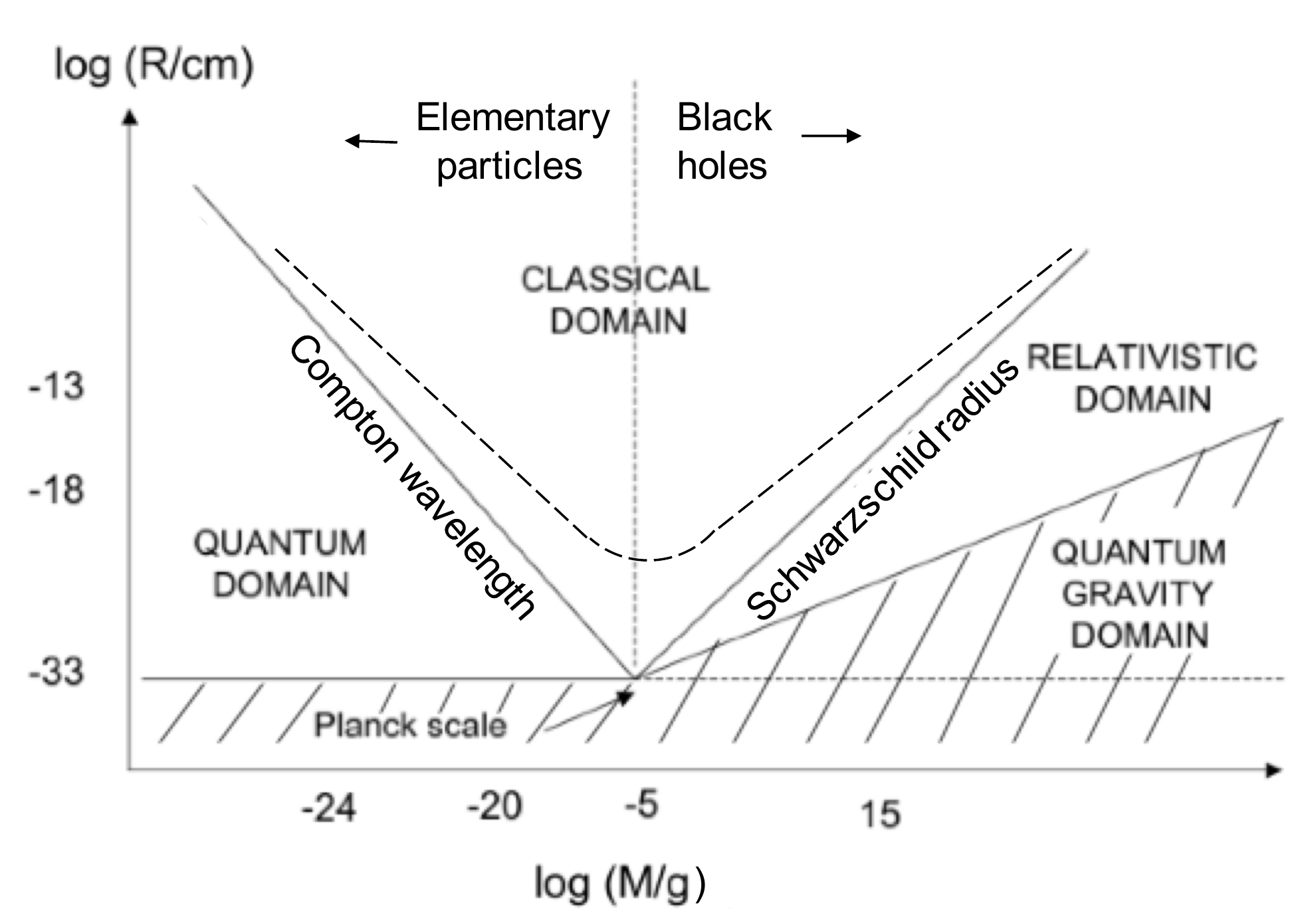}
\caption{
The division of the ($M,R$) diagram into the classical, quantum, relativistic and quantum gravity domains. The boundaries are specified by the Planck density, the Compton wavelength and the Schwarzschild radius. 
\label{MR}
}
\end{figure}

A key feature of the microscopic domain is the (reduced) Compton wavelength for a particle of rest mass $M$, which is $R_C = \hbar/(Mc)$. 
 In the $(M,R)$ diagram of Fig.~\ref{MR}, the region corresponding to $R<R_C$ might be regarded as the ``quantum domain'' in the sense that the classical description breaks down there. A key feature of the macroscopic domain is the Schwarzschild radius for a body of mass $M$, $R_S = 2GM/c^2$, which corresponds to the size of the event horizon. The region $R<R_S$ might be regarded as the ``relativistic domain'' in the sense that there is no stable classical configuration in this part of Fig.~\ref{MR}. 

The Compton and Schwarzschild lines intersect at around the Planck scales,
$R_{P} = \sqrt{ \hbar G/c^3} \sim 10^{-33} \mathrm {cm},
M_{P} = \sqrt{ \hbar c/G} \sim 10^{-5} \mathrm g$,
and divide the $(M,R)$ diagram in Fig.~\ref{MR} into three regimes, which we label quantum, relativistic and classical. 
There are several other interesting lines in the figure. The vertical line $M=M_{P}$ marks the division between elementary particles ($M <M_{P}$) and black holes ($M > M_{P}$), since the size of a black hole is usually required to be larger than the Compton wavelength associated with its mass. The horizontal line $R=R_{P}$ is significant because quantum fluctuations in the metric should become important below this \cite{wheeler}.
Quantum gravity effects should also be important whenever the density exceeds the Planck value, $\rho_{P} = c^5/(G^2  \hbar) \sim 10^{94} \mathrm {g \, cm^{-3}}$, corresponding to the sorts of curvature singularities associated with the big bang or the centres of black holes \cite{CaMoPr:2011}. This implies $R < R_{P}(M/M_{P})^{1/3}$, which is well above the $R = R_{P}$ line in Fig.~\ref{MR} for $M \gg M_P$, so one might regard the shaded region as specifying the `quantum gravity' domain. This point has recently been invoked to support the notion of Planck stars \cite{rovelli} and could have important implications for the detection of evaporating black holes \cite{barrau}.  
Note that the Compton and Schwarzschild lines transform into one another under the T-duality transformation  $M \rightarrow M_{P}^2/M$. This interchanges sub-Planckian and super-Planckian mass scales and 
 corresponds to a reflection in the line 
$M = M_{P}$ in Fig.~\ref{MR}.
T-dualities arise naturally in string theory and are known to map momentum-carrying string states to winding states and vice-versa \cite{Zwiebach:2004tj}. 

Although the Compton and Schwarzschild boundaries correspond to straight lines in the logarithmic plot of Fig.~\ref{MR}, this form presumably breaks down near the Planck point due to quantum gravity effects. One might envisage two possibilities: either there is a smooth minimum, as indicated by the broken line in Fig.~\ref{MR}, so that the Compton and Schwarzschild lines in some sense merge, or there is some form of phase transition  or critical point at the Planck scale, so that the separation between particles and black holes is maintained. Which alternative applies has important implications for the relationship between elementary particles and black holes \cite{CaMuNic:2014}.
This may also relate to the issue of T-duality since this purports to play some role in linking point particles and black holes. Such a link is also suggested by Fig.~\ref{fig:urob}.
 
One way of smoothing the transition between the Compton and Schwarzschild lines is to invoke
some connection between the Uncertainty Principle on microscopic scales and black holes on macroscopic scales. This is termed the Black Hole Uncertainty Principle (BHUP) correspondence \cite{Ca:2014a} and also the Compton-Schwarzschild correspondence when discussing an interpretation in terms of extended de Broglie relations \cite{Lake:2015pma}. It is manifested in a unified expression for the Compton wavelength and Schwarzschild radius. 
The simplest expression of this kind would be 
\begin{equation} \label{BHUP}
R_{CS} =  \frac{\beta \hbar}{Mc} + \frac{2GM}{c^2}  \, ,
\end{equation}
where $\beta$ is the (somewhat arbitrary) constant appearing in the Compton wavelength. In the sub-Planckian regime,
 this can be written as
\begin{equation} \label{GUP2}
R_C'  = \frac{\beta \hbar}{Mc} \left[1  + \frac{2}{\beta}  \left( \frac{M}{M_P} \right)^2 \right]  \quad (M \ll M_P)  \, ,
 \end{equation}
with the second term corresponding to a small correction of the kind invoked by the Generalised Uncertainty Principle  \cite{Adler1}. In the super-Planckian regime,
 it becomes
\begin{equation} \label{GEH1} 
R_S' = \frac{2GM}{c^2} \left[ 1 + \frac{\beta}{2} \left(\frac{ M_P}{ M} \right)^2 \right] \quad (M \gg M_P) \, .
\end{equation}
This is termed the Generalised Event Horizon \cite{Ca:2014a}, with the second term corresponding to a small correction to the usual Schwarzschild expression. 
More generally, the BHUP correspondence might allow any unified expression $R_C'(M) \equiv R_S'(M)$ which has the asymptotic behaviour $\beta \hbar/(Mc)$ for $M \ll M_{P}$ and $2GM/c^2$ for $M \gg M_{P}$. 
One could envisage many such expressions 
but we are particularly interested in those which -- like Eq.~(\ref{BHUP}) -- exhibit T-duality.

At the last meeting, I discussed some of the consequences of the BHUP correspondence, with particular emphasis on the  implied black hole temperature, the link with Loop Quantum Gravity \cite{CaMoPr:2011} and the effect of extra dimensions \cite{Ca:2013,koppel}. The implication is that in some sense elementary particles are  sub-Planckian black holes. Next I discuss some developments arising out of recent work with my collaborators.

\section{Carr-Mureika-Nicolini work }

The results of Ref.~\cite{CaMuNic:2014} are now summarised. In the standard picture, the mass in the Schwarzschild solution is obtained by 
matching the metric coefficients with the Newtonian potential 
and this gives the Komar integral 
\begin{equation}
M\equiv\frac{1}{4\pi G}\int_{\partial\Sigma}d^2x\sqrt{\gamma^{(2)}}\ n_\mu\sigma_\nu\nabla^\mu K^\nu \, ,
\label{komar}
\end{equation}
where $K^\nu$ is a timelike vector, $\Sigma$ is a spacelike surface with unit normal $n^\mu$, and $\partial\Sigma$ is the boundary of $\Sigma$ (typically a 2-sphere at spatial infinity) with metric $\gamma^{(2)ij}$ and outward normal $\sigma^\mu$. 
For $M\gg\Mpl$, quantum effects are negligible and one finds the usual Schwarzschild mass. For $M<\Mpl$, however, the expression can simultaneously refer to a particle and a black hole. 
One usually 
considers the particle case 
and writes Eq.~(\ref{komar}) as   
\begin{equation}
M\equiv \int_{\Sigma} d^3x \sqrt{\gamma}\ n_\mu K_\nu T^{\mu\nu}\simeq -4\pi\int_0^{\RC} dr \, r^2  T^{\ 0}_{ 0} \, ,
\label{komarparticle}
\end{equation}
where $\gamma$ is the determinant of the spatially induced metric $\gamma^{ij}$, $T^{\mu \nu}$ is the stress-energy tensor and $T^{\ 0}_{ 0}$ accounts for the particle distribution on a scale of order $\RC$. This corresponds to the mass appearing in the expression for the Compton wavelength.
When the black hole reaches the
 final stages of evaporation, 
the major contribution to integral (\ref{komar}) becomes
\begin{equation}
M = -4\pi\int_0^{\ellp} dr \, r^2  T^{\ 0}_{ 0} \, ,
\label{qbhkomar}
\end{equation}
where $T^{\ 0}_{ 0}$ accounts for an unspecified quantum-mechanical distribution of matter and energy. 
Integral (\ref{qbhkomar}) is then unknown and might lead to a completely different definition of the Komar energy. 

Inspired by the dual role of $M$ in the GUP, we explore a variant of the last
scenario, based on the existence of sub-Planckian black holes, \textit{i.e.} quantum mechanical objects  that are simultaneously black holes and  elementary particles.
n this context, we suggest that the Arnowitt-Deser-Misner (ADM) mass, which coincides with the Komar mass in the stationary case, should be 
\begin{equation}
\Madm = M \left(1+\frac{\beta}{2} \frac{\Mpl^2}{M^2}\right) \, ,
\label{newmass}
\end{equation}
which is equivalent to Eq.~(\ref{GEH1}). 
We thus posit a quantum-corrected Schwarzschild metric, like the usual one but with $M$ replaced by $\Madm$.  
We note a possible connection with the energy-dependent metric proposed in the framework of ``gravity's rainbow'' \cite{rainbow}. It may also relate to
the distinction between the bare and renormalized mass in QFT in presence of stochastic metric fluctuations \cite{camacho}.

\begin{figure}[b]
\sidecaption
\includegraphics[scale=.2]{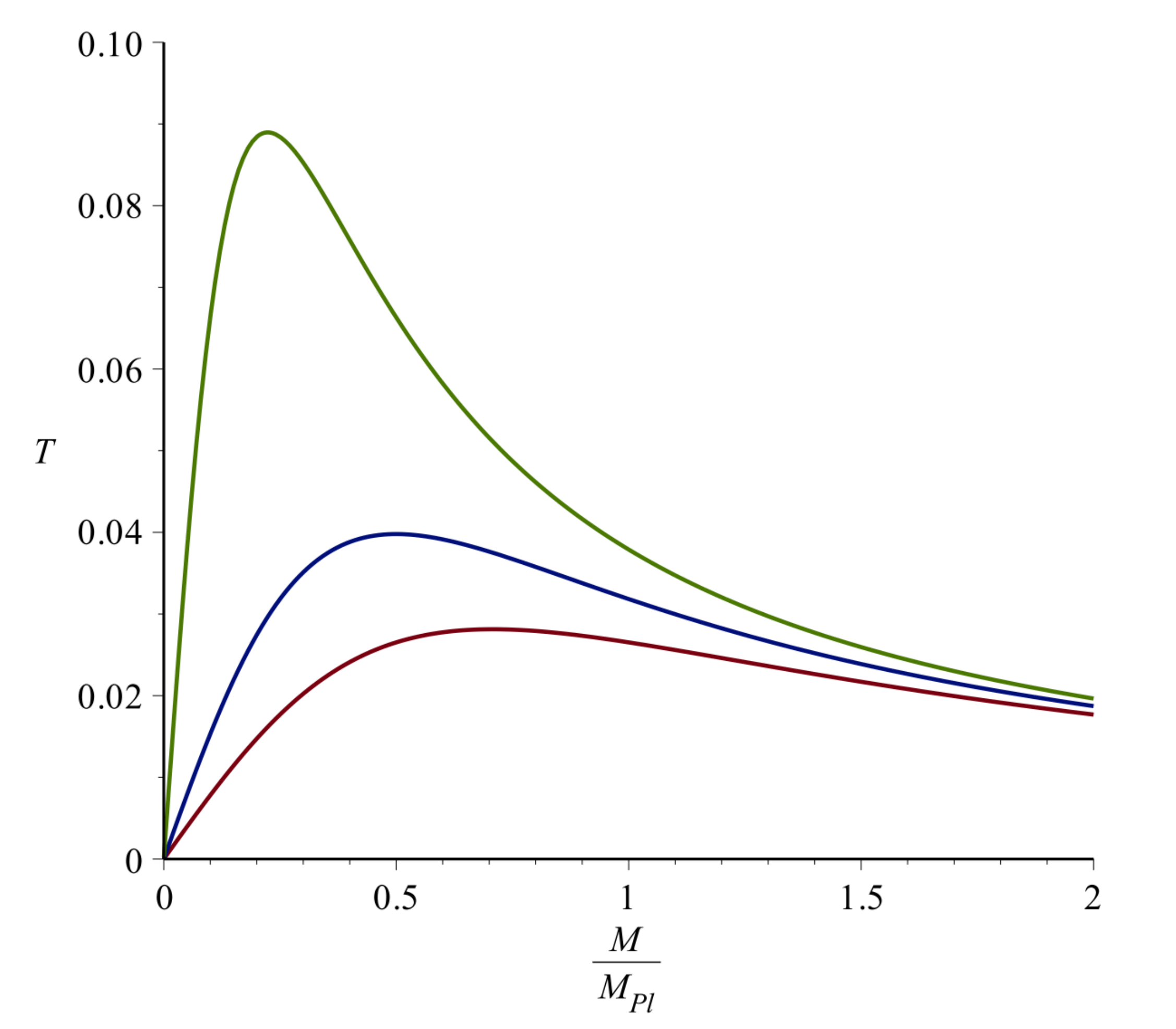}\caption{
 Hawking temperature (\ref{hawktemp}) implied by the surface gravity argument as a function of $M/\Mpl$ for $\beta = 1$ (bottom), $\beta = 0.5$ (middle) and $\beta = 0.1$ (top).  As $M$ decreases, $T$ reaches a maximum below $T_P$ and then falls to zero. 
\label{fig1}
}
\end{figure}

The horizon size for the modified metric
is given by
\bea
R_S' = \frac{2 \Madm} {\Mpl^2} 
 \approx
\left\lbrace
\begin{array}{ll}
2M/  \Mpl^2 & (M \gg \Mpl ) \\
(2+\beta)/\Mpl   & (M \approx \Mpl ) \\
\beta/M &(M \ll \Mpl ) \, ,
\end{array}
\right.
\label{horizon}
\eea
The first expression is 
the standard Schwarzschild radius.  The intermediate expression gives a minimum 
of order $\ellp$, so the Planck scale is never actually reached for $\beta>0$ and the singularity remains inaccessible.
The last expression 
resembles the Compton wavelength. If the temperature is determined by the black hole's surface gravity \cite{hawking}, one has
\bea
T  = \frac{\Mpl^2}{8\pi \Madm } \approx
\left\lbrace
\begin{array}{ll}
\Mpl^2/(8\pi M) [ 1- \beta  (\Mpl / M)^{2} ]  & (M \gg \Mpl) \\
\Mpl /(8\pi(1+\beta/2))  & (M \approx \Mpl) \\
M/(4\pi \beta) [ 1- (M/ \Mpl)^{2}/\beta ] & (M \ll \Mpl)  \, .
\end{array}
\right.
\label{hawktemp}
\eea
This temperature is plotted in Fig.~\ref{fig1}.
The large $M$ limit  is the usual Hawking temperature with a small correction.
However, as the black hole evaporates, the temperature reaches a maximum at around $T_P$
and then decreases to zero 
as $M \rightarrow 0$. 

A possible explanation for the $M \ll M_{Pl}$ behaviour is that a decaying black hole makes a temporary transition to a (1+1)-D dilaton black hole when approaching the Planck scale,
since  this naturally encodes a $1/M$ term in its gravitational radius.
For according to t'Hooft \cite{thooft}, gravity might experience a (1+1)-D phase at the Planck scale due to spontaneous dimensional reduction, such a conjecture being further supported by studies of the fractal properties of a quantum spacetime at the Planck scale. 
At this point  the Komar mass can be defined  as for dilaton black holes by \cite{jmpnprd}
\begin{equation}
M\sim \int dx\sqrt{g^{(1)}} \, n_i ^{(2)}T^{\ i}_{ 0} \, ,
\label{qbhkomar2d}
\end{equation}
where $g^{(1)}$ is the determinant of the spatial section of $g^{ij}$, the effective 2D quantum spacetime metric, and $^{(2)}T^{\ i}_{ 0}$ is the dimensionally reduced energy-momentum tensor. 

The black hole luminosity 
in this model is
$L = \gamma^{-1} \Madm^{-2}$
where $\gamma \sim t_\mathrm{Pl}/\Mpl^3$. Although the black hole loses mass on a timescale 
$\tau  \sim M/L \sim \gamma M^{3}(1+ \beta \Mpl^2/2M^2)^{2}$,
it never evaporates entirely because the mass loss rate decreases when $M$ falls below $M_P$. There are two values of $M$ for which $\tau$ is comparable to the age of the Universe ($t_0 \sim 10^{17}$s).
One is super-Planckian,
$M_* \sim (t_0/ \gamma)^{1/3} \sim (t_0/\tpl)^{1/3} \Mpl  \sim 10^{15}$g,
this being the standard expression for the mass of a PBH evaporating at the present epoch, and the other is sub-Planckian, 
$M_{**} \sim  \beta^2 (\tpl/t_o) \Mpl  \sim 10^{-65}$g.
The usual Hawking lifetime ($\tau \propto M^3$) gives the time for this mass to decrease to $\Mpl$, after which it quickly falls to the value $M_{**}$. 
Although this mass-scale is very tiny, it arises naturally in some estimates for the photon or graviton mass 
\cite{rbmjm}). 
Note that the PBH mass cannot actually reach $M_{**}$ at the present epoch because
 the black hole temperature is less than the CMB temperature
below $M_{\rm CMB} \sim 10^{-36}$g, 
leading to {\it effectively} stable relics which might provide the dark matter.
 
To summarise the advantages of our proposal: it encodes the GUP duality in the expression for the mass; it smooths the $M(R)$ curve, so that there is no critical point;  it cures the thermodynamic instability of evaporating black holes; it exhibits dimensional reduction in the sub-Planckian regime;  and it gives a consistent theory of gravity in different spacetime dimensions without needing two regimes governed by different theories (GR and QM). Indeed, in some sense, the BHUP correspondence implies that  all black holes are quantum and that the Uncertainty Principle has a gravitational explanation.

\section{Lake-Carr work}

Canonical (non-gravitational) quantum mechanics is based on the concept of wave-particle duality, encapsulated in the de Broglie relations $E = \hbar\omega$ and $p = \hbar k$. When combined with the 
energy-momentum relation for a non-relativistic point particle, these
lead to the dispersion relation $\omega = (\hbar/2m)k^2$. 
However, these relations break down near the Planck scale, since they correspond to wavelengths $\lambda \ll R_P$ or periods $t \ll t_P$. Ref.~\cite{Lake:2015pma} therefore proposes modified forms for the  de Broglie relations  which may be applied even for $E \gg M_Pc^2$, 
with the additional terms being interpreted as representing the self-gravitation of the wave packet. 

\begin{figure}[b]
\sidecaption
\includegraphics[scale=.5]{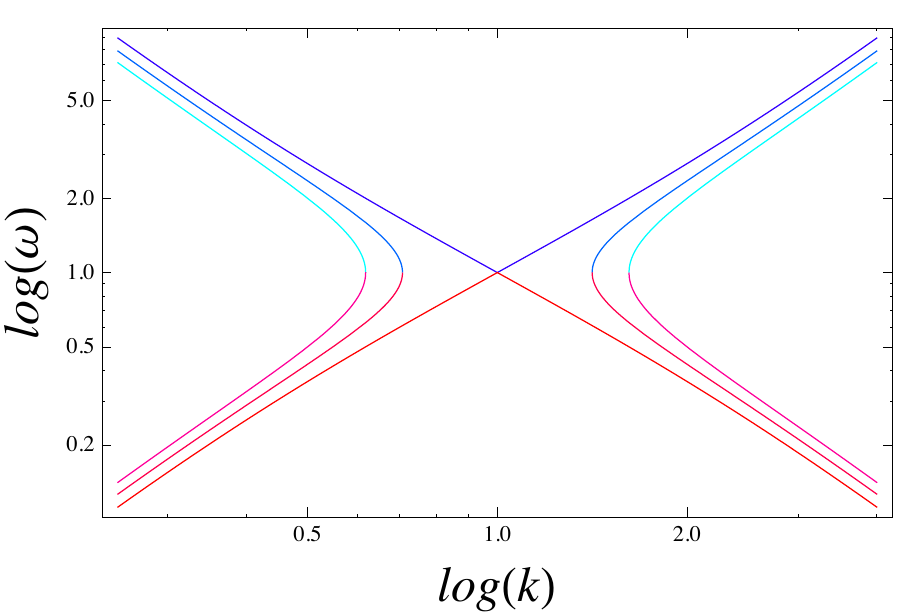}
\includegraphics[scale=.5]{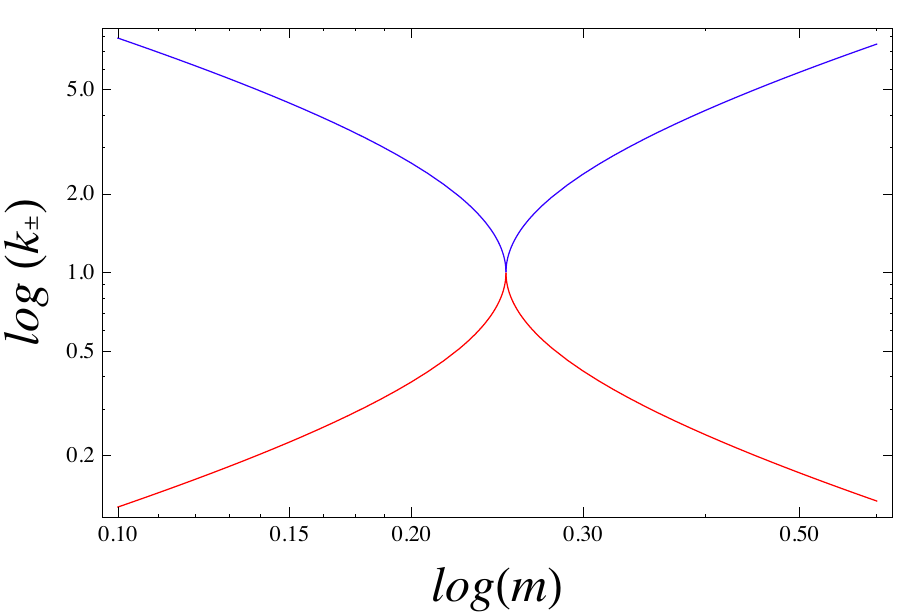}
\caption{
Illustrating how (a) the dispersion relations $\omega_{\pm}(m,k)$ for three values of $m$ and (b) the limiting wavenumbers $k_{\pm} (m)$ are changed  in the proposed model.
\label{fig4}
}
\end{figure}

The simplest such relations are
$E = \hbar\Omega$ and $p = \hbar \kappa$ 
with
\begin{equation} \label{Omega1}
\Omega = \left \lbrace
\begin{array}{rl}
\omega_P^2\left(\omega + \omega_P^2/\omega\right)^{-1} & \ (m < M_P) \\
\beta\left(\omega + \omega_P^2/\omega\right) & \ (m > M_P) \ \ \
\end{array}\right.
, \, \kappa = \left \lbrace
\begin{array}{rl}
k_P^2\left(k + k_P^2/k\right)^{-1} & \ (m < M_P) \\
\beta\left(k + k_P^2/k\right) & \ (m > M_P).
\end{array}\right.
\end{equation}
Continuity of $E$, $p$, $dE/d\omega$ and $dp/dk$ at $\omega = \omega_P$ and $k=k_P$ 
is ensured by setting $\beta = 1/4$.  
The relation  $\Omega = (\hbar/2m)\kappa^2$ then leads to new dispersion relations, quadratic in $\omega$,  which  can be solved for both $E \ll M_pc^2$ and $E \gg M_pc^2$. 
The two solution branches, $\omega_{\pm}(k,m)$, 
are shown as functions of $k$ for 
the three values of $m$
 in Fig.~\ref{fig4}(a). 
The solutions are dual under the transformation $m \rightarrow M_P'^2/m$ where $M_P' \equiv (\pi/2)M_P$. 
Canonical non-relativistic quantum mechanics is recovered in the bottom left region,
where $\omega_{-} \approx (\hbar/2m)k^2$. The branches meet at $\omega_{\pm}(k_P) = \omega_P$ for the critical case $m = M_P'$ but there is a gap in the allowed values of $k$ 
for $m \neq M_P'$. 
The limiting values 
for a given mass, $k_{\pm}(m)$, are shown in Fig.~\ref{fig4}(b) and these also exhibit duality.
These values correspond to the Schwarzschild formula for $E \gg M_Pc^2$ and the Compton formula for $E \ll M_Pc^2$. So this is another way of interpreting the BHUP correspondence. 

\begin{figure}[b]
\sidecaption
\includegraphics[scale=.13]{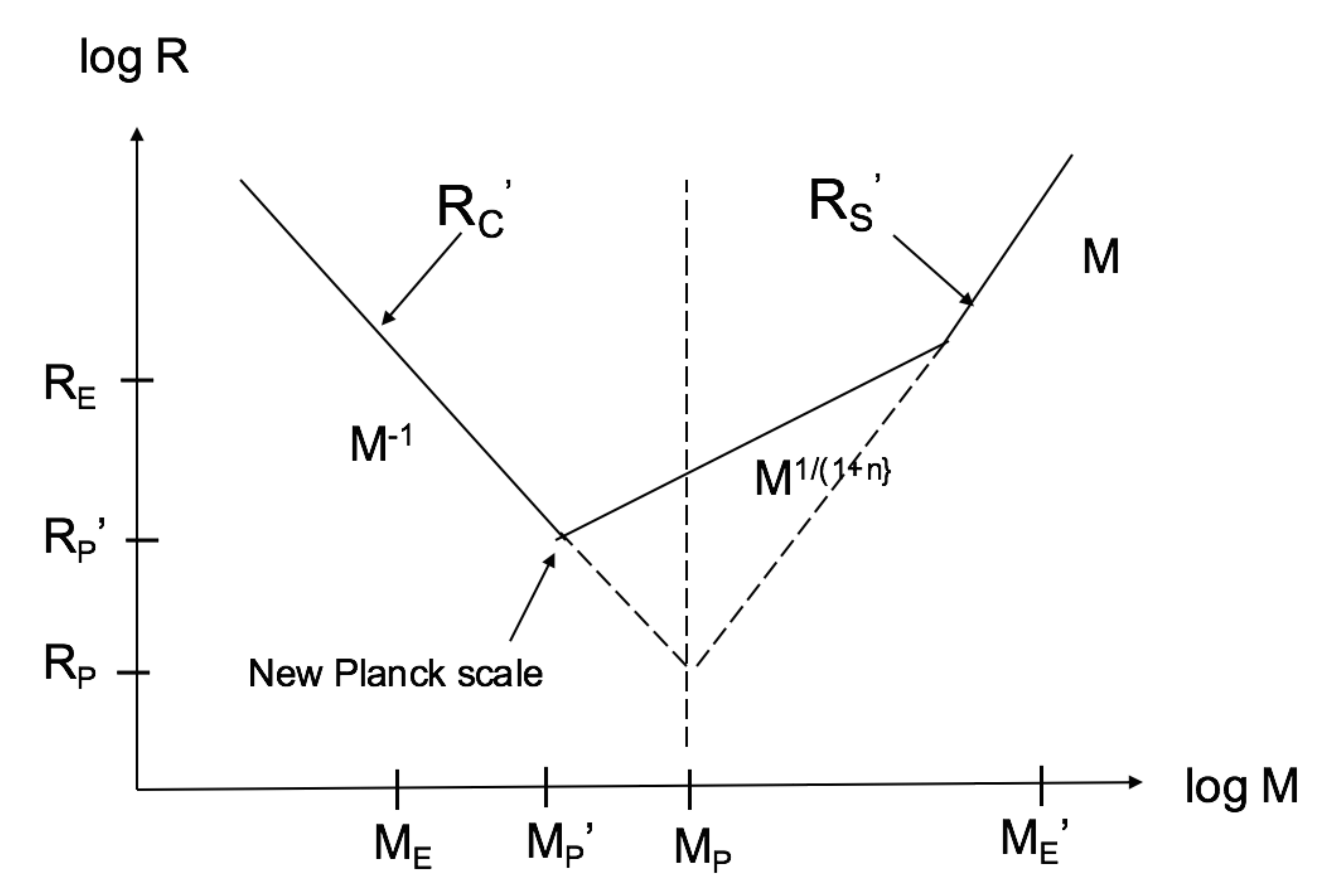}
\includegraphics[scale=.25]{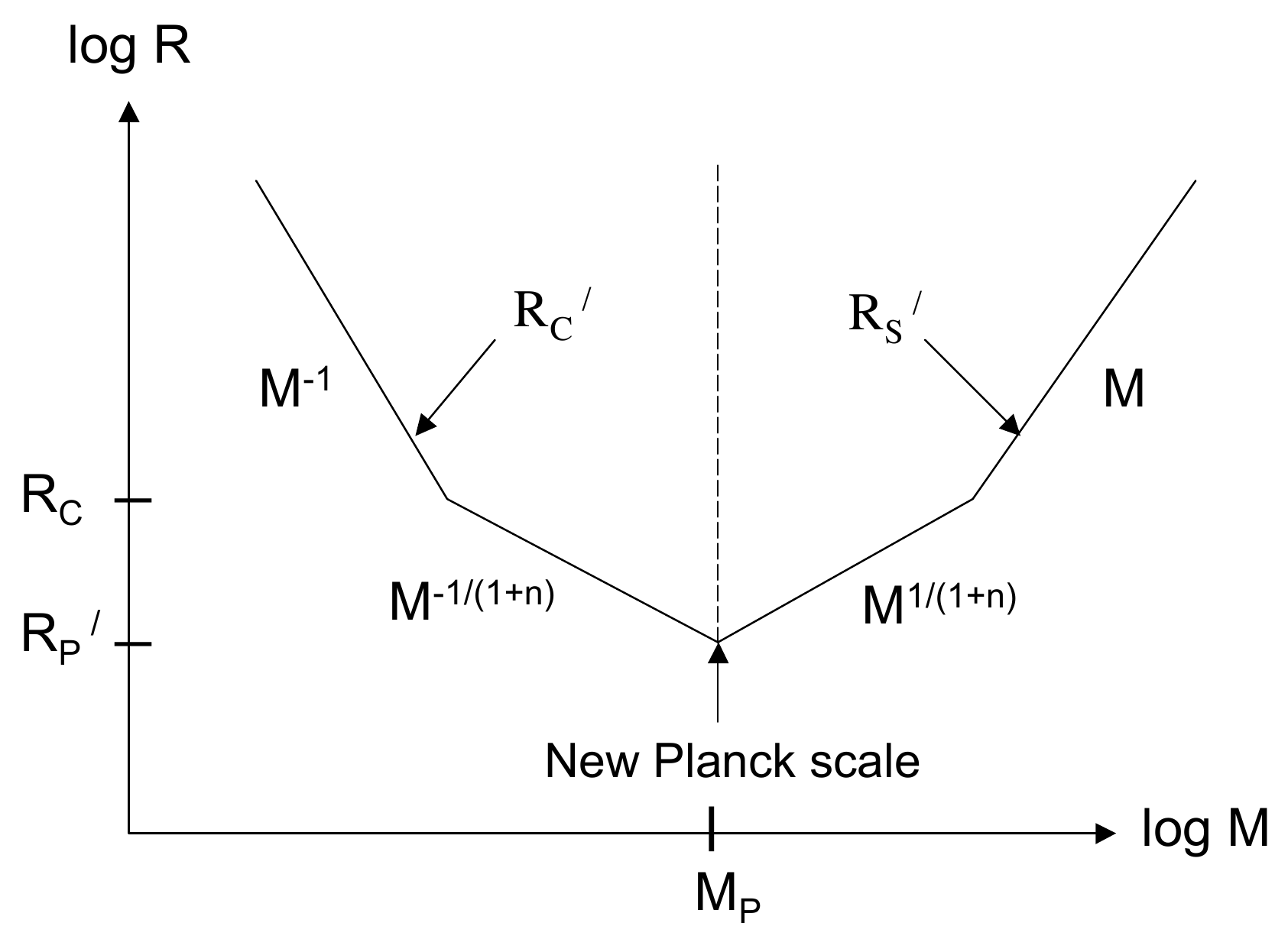}
\caption{
Showing change in Planck scales for large extra dimensions if (a) only the Schwarzschild radius is modified and (b) the Compotn wavelegnth is also modified,  preserving T-duality. 
\label{fig5}
}
\end{figure}

In our second paper \cite{lake} we discuss the preservation of T-duality in higher dimensions. In three spatial dimensions, the Compton wavelength and Schwarzschild radius  are dual under the transformation $M \rightarrow M_{P}^2/M$.
In the presence of $n$ extra dimensions, compactified on some scale $R_E$, it is usually assumed that $R_S \propto M^{1/(1+n)}$ \cite{kanti2004} and $R_C \propto M^{-1}$ (as in three domensions) for $R < R_E$, which breaks the duality. This situation is illustrated in Fig.~\ref{fig5}(a) and gives the standard scenario in which the effective Planck length is increased and the Planck mass reduced, allowing the possibility of 
black hole production at the LHC.  

Currently there is no evidence for such production. However, 
 the effective Compton wavelength depends on the form of the $(3+n)$-dimensional wavefunction.  If this is spherically symmetric, then one indeed has $R_C \propto M^{-1}$. But if the wave function is pancaked in the extra dimensions and maximally asymmetric, then $R_C \propto M^{-1/(1+n)}$, so that the duality between $R_C$ and $R_S$ is preserved. This situation is illustrated in Fig.~\ref{fig5}(b), which shows that the effective Planck length is reduced even more but the Planck mass is unchanged. So TeV quantum gravity is precluded in this case and black holes {\it cannot} be generated in collider experiments. Nevertheless, the extra dimensions could still have consequences for the detectability of black hole evaporations and the enhancement of pair-production at accelerators on scales below $R_E$. 

\begin{acknowledgement}
I thank my collaborators in the work reported here: Matthew Lake, Jonas Mureika and Piero Nicolini.
\end{acknowledgement}

\end{document}